\documentclass[a4paper,fleqn,usenatbib]{mnras}

\usepackage{newtxtext,newtxmath}

\usepackage[T1]{fontenc}
\usepackage{ae,aecompl}

\usepackage{graphicx}	
\usepackage{amsmath}	
\usepackage{tablefootnote}

\newcommand{\fesc}{$f_{\rm esc}$}
\newcommand{\fpdf}{$f_{\rm esc}^{\rm PDF}$}
\newcommand{\fTm}{$f_{\rm esc}^{\langle T \rangle}$}
\newcommand{\fbias}{$f_{\rm esc}^{\rm bias}$}
\newcommand{\Tm}{$\langle T_{\rm IGM} \rangle$}
\newcommand{\Lint}{($L_{900}/L_{1500}$)$_{\rm int}$}
\newcommand{\TIGM}{$T_{\rm IGM}$}
\newcommand{\Tbias}{$T_{\rm bias}$}
\usepackage{mathtools}

\title[A Probabilistic Estimate of {\fesc}]{A Cautionary Tale of
 LyC Escape Fraction Estimates from High Redshift Galaxies}

\author[R. Bassett et al.]{
R. Bassett$^{1,2}$\thanks{E-mail: rbassett.atsro@gmail.com (RB)},
E. V. Ryan-Weber$^{1,2}$\thanks{E-mail: eryanweber@swin.edu.au (ERW)},
J. Cooke$^{1,2}$,
U. Me\v{s}tri\'{c}$^{1,2}$,
L.J. Prichard$^{3}$,\newauthor
M. Rafelski$^{3,4}$,
I. Iwata$^{5}$, 
M. Sawicki$^{6}$\thanks{Canada Research Chair}, 
S. Gwyn$^{7}$,
S. Arnouts$^{8}$
\\
$^{1}$Centre for Astrophysics and Supercomputing, Swinburne University
of Technology, PO Box 218, Hawthorn VIC 3122, Australia\\
$^{2}$ARC Centre of Excellence for All Sky Astrophysics in 3 Dimensions (ASTRO 3D), Australia\\
$^{3}$Space Telescope Science Institute, 3700 San Martin Drive,
Baltimore MD 21218, USA\\
$^{4}$Department of Physics \& Astronomy, John Hopkins University,
Baltimore, MD 21218, USA\\
$^{5}$National Astronomical Observatory of Japan, 2-21-1 Osawa,
Mikata, Tokyo 181-8588, Japan\\
$^{6}$Department of Astronomy \& Physics and the Institute for
Computational Astrophysics, Saint Mary's University,\\ 923 Robie Street,
Halifax, Nova Scotia, B3H 3C3, Canada\\
$^{7}$NRC-Hertzberg, 5071 West Saanich Road, Victoria, British
Columbia, V9E 2E7, Canada\\
$^{8}$Aix Marseille Université, CNRS, LAM - Laboratoire
d’Astrophysique de Marseille, 38 rue F. Joliot-Curie, F-13388,
Marseille, France 
}

\date{Accepted XXX. Received YYY; in original form ZZZ}

\pubyear{2021}

\begin{document}
\label{firstpage}
\pagerange{\pageref{firstpage}--\pageref{lastpage}}
\maketitle

\begin{abstract}
Measuring the escape fraction, {\fesc}, of ionizing, Lyman Continuum
(LyC) radiation is key to our understanding of the process of cosmic
reionization. In this paper we provide a methodology for
recovering the posterior probability distribution of the LyC escape
fraction, {\fpdf}, considering both the observational uncertainties and
ensembles of simulated transmission functions through the
intergalactic medium (IGM). We present an example of this method
applied to a VUDS galaxy at $z=3.64$ and find {\fpdf} =
0.51$^{+0.33}_{-0.34}$ and compare this to the values computed assuming
averaged IGM transmission with and without consideration of detection
bias along average sightlines yielding
{\fTm} = 1.40$^{+0.80}_{-0.42}$, 
and {\fbias} = 0.82$^{+0.33}_{-0.16}$. Our results highlight the
limitations of methods assuming average, smooth transmission
functions. We also present MOSFIRE data for a sample of seven LyC
candidates selected based on photometric redshifts at $z > 3.4$, but
find that all seven have overestimated photometric redshifts by
$\Delta z \sim 0.2$ making
them unsuitable for LyC measurements. This results likely due to a
bias induced by our selection criteria. 
\end{abstract}

\begin{keywords}
intergalactic medium -- galaxies: ISM -- dark ages, reionization,
first stars
\end{keywords}



\section{Introduction}

The epoch of reionization (EoR), the period during which the hydrogen
permeating the intergalactic medium (IGM) was photoionized by young
galaxies, is currently an extremely active area of research covering a
wide range of topics and methods
\citep[e.g.][]{ghara21,hutter21,pagano21}. Although our understanding of the timeline of
reionization is continually being refined \citep[e.g.][]{bolton07,robertson15,adam16}, there remain
a number of key, outstanding questions, such as what are the primary
drivers of the reionization process? Inevitably, the ionizing,
Lyman Continuum (LyC) photons responsible originate in galaxies:
either from stellar sources \citep[e.g. massive O and B stars and
X-ray binaries in star-forming galaxies][]{eldridge17,shivaei18} or from active galactic nuclei
\citep[AGN, e.g.][]{grazian18}. Currently star-forming galaxies are favoured while
AGN are expected to play a larger role in sustaining the UV background
radiation that maintains an ionized IGM at lower redshifts \citep{kakiichi18}.

The difficulty in definitively answering the question of which sources
are primarily responsible for reionization can be attributed
to the faintness of these sources \citep[e.g.][]{bian20,mestric20} and the
high opacity of the IGM during the EoR. Based on
known samples of Lyman Break Galaxies \citep[LBGs]{steidel18} and Lyman $\alpha$
emitters \citep[LAEs][]{fletcher18}, \citet{bassett21} predict that
LyC emission from the bulk of star-forming galaxies at $z \geq 3.0$
should be fainter than 28 mag AB. The low flux of ionizing photons
reaching the Earth is predominantly the result of absorption by
hydrogen both within the host galaxy's interstellar medium (ISM), the
circumgalactic medium (CGM), and
the intervening IGM. The ISM/CGM absorption is directly related to the escape fraction of LyC photons
{\fesc}, the measurement of which is the primary goal of a number of
observational programs \citep[e.g.][]{marchi17,wang21}. Variations in {\fesc} with
other galaxy properties such as mass \citep[e.g.][]{naidu19}, size and morphology
\citep[e.g.][]{kim21}, or time varying star-formation rate \citep[SFR,][]{smith19}  can
have significant implications regarding how the reionization process
proceeds. 

In order to measure such dependence on {\fesc} and galaxy
properties, we must first be able to accurately estimate {\fesc} from
observations. The largest difficulty in achieving this goal is the
highly stochastic transmission of the IGM to ionizing photons
\citep[e.g.][]{inoue08,steidel18,bassett21} as this quantity is
degenerate with the value of {\fesc} inferred from observations. One
strategy is to assume the average
IGM transmission across a large ensemble of simulated transmission
curves \citep[e.g.][]{inoue14}, although this technique is likely to be
appropriate only for statistically significant samples of LyC
detections, which are currently lacking. One can also apply each IGM
transmission curve from an ensemble individually and provide a
histogram of the resulting {\fesc} values
\citep[e.g.][]{shapley16,vanzella16}, though this alone provides poor
constraint on {\fesc} and produces a large number of sightlines with
{\fesc} > 1 (i.e. more LyC photons escape the galaxy than are expected
to be intrinsically produced). Future telescopes and instruments \citep[e.g. the Keck
Wide-Field Imager][]{gillingham20} are expected to push LyC observations to
greater depths, thus we may be on the cusp of the era of large LyC
samples. In light of this, it is important to reassess the methodology
of estimating {\fesc} from observed galaxies.

In this paper we aim to provide a statistical framework for
determining the posterior probability distribution function (PDF) for
{\fesc} from individual LyC detected galaxies. This method applies an
ensemble of 10,000 IGM 
transmission curves and tests 10,000 {\fesc} values for each of these
possible sightlines, measuring the resulting LyC flux for an ensemble of
SEDs fit using a
large grid of BPASSv2.1 \citep{eldridge17} models. The modeled fluxes
are considered in a 
probabilistic manner combining both the observed LyC flux and
uncertainty as well as the goodness of fit of each BPASS model to 20
photometric bands at $\lambda_{\rm rest} > 1216 {\AA}$ (thus avoiding
light attenuated by the IGM) producing a single {\fesc} PDF. The paper
is organised as follows: in Section \ref{section:sample} we outline
the sample selection and observations, in Section \ref{section:method}
we discuss the methodology including redshift measurement, providing a
discussion of systematic overestimates of photometric redshifts
(likely resulting from our selection criteria) then focus {\fesc}
estimates for a single galaxy at $z > 3.4$, in Section
\ref{section:results} we present the results of
our analysis for this galaxy, and in Section
\ref{section:conclusions} we summarise our findings.

\section{Sample Selection and Observations}\label{section:sample}

All but one of the LyC emitting candidate galaxies in our preliminary
sample were selected from the
ZFOURGE survey \citep[PI Labbe,][]{straatman16} for spectroscopic observations with the
Multi-Object Spectrometer For Infra-Red Exploration 
\citep[MOSFIRE][]{mclean12}. The ZFOURGE survey provides robust
photometric redshifts by utilising 30+ photometric bands from $u^{*}$-band
to far-IR. The ZFOURGE team has estimated an average photometric redshift
accuracy of $\pm$2\% based on subsamples of galaxies with
spectroscopic follow-up observations. It is relevant to point out that
the photometric redshift accuracy quoted by ZFOURGE is primarily based
on galaxies at $z < 3$ where the split J and H band filters probe the
Balmer break directly. At higher redshifts, the Balmer break moves
into the K band, thus the photometric redshifts are likely to become
less reliable (see also Appendix \ref{section:zoverest}). The final galaxy in our sample
is selected from the VIMOS Ultra Deep Survey \citep[VUDS][ID
511227001]{marchi15}, and already has a secure spectroscopic redshift
measurement of $z=3.64$. We carried out spectroscopic observations
using MOSFIRE to confirm ZFOURGE photometric redshifts for the remaining
targets (see Section \ref{section:lineflux}). 

\begin{figure*}
\includegraphics[width=\textwidth]{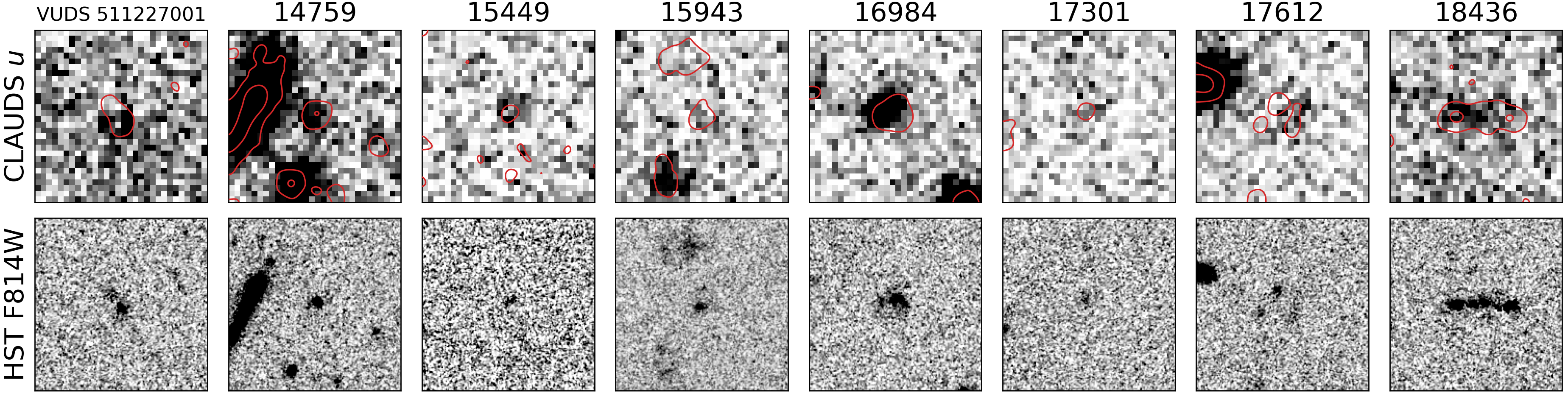}
  \caption{Imaging data used for sample selection with a field-of-view
    of 5x5 arcsec per cutout. \textit{Top:} CLAUDS $u$ band imaging for our sample. Red
    contours are taken from the HST F814W imaging after smoothing with
  a Gaussian kernel with a width of 4 pixels. The purpose here is to
  illustrate the relative position of the source in each
  image. \textit{Bottom:} HST F814W imaging of our sample with the
  same centering and field-of-view as for the top row.}
  \label{fig:HST}
\end{figure*}

Accurate redshifts are critical as our key selection criteria for LyC emitting
candidates is clean $u$-band detection \citep[see, e.g.][]{bassett19}, probing LyC flux above
$z\sim3$ (exclusively so above $z=3.4$). As in previous works
\citep[e.g.][]{bassett19,mestric20} $u$-band data comes from
the CFHT Large Area U-band Deep Survey \citep[CLAUDS,][]{sawicki19}, which
reaches a maximum depth of $\sim$27.2-27.3 mag in the $u$-band. We
note that the $u$-band used here for LyC detections is distinct
from the $u^{*}$-band of the ZFOURGE survey. In particular the
$u$-band for the CLAUDS survey exhibits a sharp cutoff in transmission
on the red end and does not suffer from red leak, a major drawback of
the $u^{*}$ band for LyC studies. 

All targets have available, multiband HST photometric data and for each we have
performed visual comparison between HST F814W and the CLAUDS $u$-band
imaging to exclude targets with a likely companions in the
space-based imaging not apparent from the ground. The criteria for excluding
galaxies based on this visual inspection are the presence of either multiple
HST detections associated with a single $u$-band detection or an offset between the
$u$-band centroid and the F814W centroid larger than the HST PSF. From our experience performing
$u$-band selections in the COSMOS field, we find roughly one third of galaxies
exhibit close pairs in higher resolution imaging \citep[see][]{mestric20}. Comparison between
$u$-band and F814W imaging is shown in Figure \ref{fig:HST} with
contours in the upper panel illustrating the smoothed HST
photometry. The field-of-view for each target is 5x5 square arcsec. This step provides
more confidence that many of our LyC emitting candidate galaxies do not have their
$u$-band flux contaminated by lower redshift interlopers
\citep[e.g.][]{vanzella10}. However, some targets with multiple HST peaks
have been included due to the limited field of view observable in a
single MOSFIRE mask. Thus, our sample represents a set aimed to optimise a
single MOSFIRE mask in the ZFOURGE-COSMOS field. More details of our $u$-band selection
can be found in \citet{mestric20}.  

LyC emitting galaxy candidates considered in this work have been
observed using the MOSFIRE instrument
at the Keck Observatory (proposal ID 2018B\_W151 PI Bassett). 
MOSFIRE observations were performed in two half nights of December
2018 in the same manner as for galaxies described in \citet{bassett19}. We observed
in the H and K bands with K
band observations targeting [OIII] ($\lambda$5007 and $\lambda$4959 {\AA}) and
H$\beta$, and H band observations targeting [OII]
($\lambda\lambda$3727 {\AA}). In both cases we employed a 1$\farcs$0
slit and a ABBA dither pattern with a 1$\farcs$25 nod. K-band observations
were performed on December 17th with 60$\times$180s exposures, a
total of 3 hours on source. The seeing for our K-band observations
varied from $\sim$0$\farcs$5 to $\sim$0$\farcs$61. 
H-band data were collected on December 19th using
100$\times$120s exposures, a total of 3.3 hours on source. Seeing
conditions for our H-band observations were similar to those for
K-band. We reduced the data using a combination of the standard
MOSFIRE python reduction
package\footnote{https://keck-datareductionpipelines.github.io/MosfireDRP/}
as well as custom scripts for flux calibration. For more details on
our data reduction process, see \citet{bassett19}. Our final sample
consist of seven ZFOURGE and one VUDS galaxy observed with MOSFIRE. 

\section{Method}\label{section:method}

\subsection{MOSFIRE Analysis}\label{section:lineflux}

\begin{figure}
\includegraphics[width=\columnwidth]{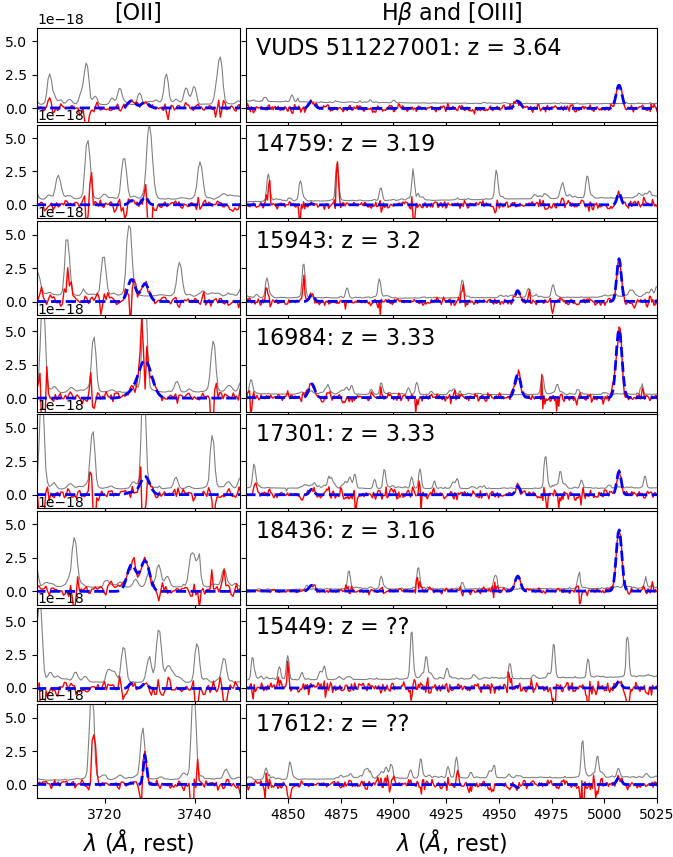}
  \caption{Fits to MOSFIRE observations of  [OII]
    $\lambda\lambda$3727, H$\beta$, [OIII] $\lambda$4959, and [OIII]
    $\lambda$5007 for our sample. The observed spectra and 1$\sigma$
    uncertainties are shown in red and grey, respectively, and our best fit
    model is shown in blue. The primary use of these data are to
    secure spectroscopic redshifts from for subsequent analysis of
    {\fesc}. Galaxies 15449 and 17612 display marginal recovery of
    [OIII] $\lambda$5007, thus we do not claim to have reliably
    secured spectroscopic redshifts for these targets. The remaining
    galaxies all display secure detections of [OIII] $\lambda$5007, however.}
  \label{fig:linefits}
\end{figure}

\begin{table*}
  \vspace{2mm}
  \begin{center}
  \begin{tabular}{ c c c c c c c c }
  \hline\hline
  ID & RA & DEC & $z$ & f([OII]$\lambda\lambda3727$) & f(H$\beta$) & f([OIII]$\lambda4959$) & f([OIII]$\lambda5007$)\\
  \hline
  VUDS511227001 & 150.062182 & 2.423024 & 3.641 & 9.2$\pm$5.5 & 8.7$\pm$13.1 & 8.8$\pm$8.4
  & 28.1$\pm$10.4\\
  14759 & 150.0642853 & 2.3365283 & 3.189 & $<$6.6 & $<$4.2 & $<$5.4 & 8.9$\pm$8.1\\
  15449 & 150.0592499 & 2.3442955 & -- & $<$7.5 & $<$7.7 & $<$8.2 & $<$9.3\\ 
  15943 & 150.0710449 & 2.3490129 & 3.202 & $<$12.7 & 5.6$\pm$13.6 & 9.1$\pm$14.1 & 36.0$\pm$13.5\\ 
  16984 & 150.0836487 & 2.3587866 & 3.327 & $<$14.8 & 15.3$\pm$26.4 & 25.9$\pm$28.7 & 75.6$\pm$26.7\\ 
  17301 & 150.0789032 & 2.3626015 & 3.325 & $<$17.6 & $<$8.7 & $<$6.3 & 19.9$\pm$8.4\\ 
  17612 & 150.0817566 & 2.3652375 & -- & $<$4.5 & $<$5.4 & $<$7.8 & $<$9.4\\ 
  18436 & 150.0518951 & 2.3979318 & 3.156 & 44.1$\pm$11.6 & 5.8$\pm$25.4 & 15.1$\pm$24.9 & 61.6$\pm$23.6\\ 
  \hline
  \end{tabular}
  \caption{Results of our MOSFIRE analysis, all line fluxes in units of 1$\times$10$^{-18}$ ergs s$^{-1}$ cm$^{-2}$.}\label{table:mosfire}
  \end{center}
\end{table*}

Our method for measuring emission line fluxes is the same as decribed in Section
4.1 of \citet{bassett19}. We simultaneously fit all emission
lines using Gaussian profiles where the $\sigma$ of all profiles are
assumed to be the same. In this way, we are able to fit for the galaxy
redshift directly rather than fitting the Gaussian centroid of each emission
line separately. Thus, from our line-fitting procedure we extract the
spectroscopic redshift, average $\sigma$ for line-emitting gas, and
the integrated line fluxes of the [OII] $\lambda\lambda$3727 {\AA}
(MOSFIRE resolution provides only marginal separation of this
doublet), H$\beta$, [OIII]
$\lambda$4959 {\AA}, and [OIII] $\lambda$5007 {\AA} emission lines. 

We show the emission line fits for our sample in Figure
\ref{fig:linefits}. For six out of eight targets, we detect  [OIII]
$\lambda$4959 {\AA}, [OIII]
$\lambda$5007 {\AA}, and H$\beta$ reliably enough to determine the spectroscopic
redshift. Galaxies 15449 and 17612 provide unconvincing detections of
[OIII] $\lambda$5007 meaning we can not obtain confident
spectroscopic redshifts. These two targets are therefore omitted from
further analysis. The recovery of other lines varies from target to
target. In particular, the [OII] $\lambda \lambda$3727 doublet often
suffers from contamination from sky emission, which largely prohibits
any meaningful line flux measurement. Regardless, the primary use of our MOSFIRE
observations is to secure accurate redshifts and provide a basis for
our $u$ band detections in the context of LyC escape. 

As stated in Section \ref{section:sample}, our targets were
specifically selected at $z > 3.4$ in order to ensure that the $u$-band
detections probe LyC photons exclusively. However, Figure \ref{fig:linefits}
shows that all of the targets selected from ZFOURGE
based on photometric redshift estimates fall below this redshift
cutoff. This means that our $u$-band detections contain contamination
from Lyman $\alpha$ forest light. In general, our SED models contain
significantly larger flux at $\lambda$ $>$ 911.8 {\AA} when compared
to LyC wavelengths. Combined with the fact that we expect no
correlation between the IGM transmission shortward of the Lyman limit
and the transmission in the Lyman forest
\citep[e.g.][]{shapley06,bassett21}, we are unable
to reliably constrain {\fesc} for galaxies at $z < 3.4$. Thus, the
remaining ZFOURGE galaxies are removed from the sample for further
analysis and only VUDS 511227001 will be considered in the context of
LyC escape. 

The LyC emission for VUDS511227001 has already been explored by \citep[][using VIMOS
spectroscopy, 1.3$\sigma$]{marchi17} and
\citet{mestric20}. The latter study makes use of the same CLAUDS $u$
band observations used here in which VUDS 511227001 is detected at
27.82 mag (3.05$\sigma$). Although the $u$-band probes bluer wavelengths
than those probed by \citet{marchi17}, where LyC emission is expected to be weaker
due to increased IGM attenuation, the increased depth of CLAUDS
when compared to VIMOS mean these two observations are consistent.
We estimate a limiting magnitude from the non-detection of \citet{marchi17} 
(based on the conversion of the 1$\sigma$ error of flux density) to be
$\sim$26.6 mag, significantly brighter than the photometric detection presented here and
in \cite{mestric20}, suggesting that the true brightness of VUDS511227001
was beyond the limit of VUDS.

\subsection{SED Models and Dust Attenuation}\label{section:SEDmodels}

The remainder of this section is focused on estimating {\fesc} for the
single target at $z > 3.4$, VUDS 511227001. The first step in this process is the selection of
a model spectral energy distribution (SED) to which the observed HST and $u$-band photometry will be
compared. Ultimately, the key property of any SED model in this
context is the ratio of intrinsic luminosity at 880-910 {\AA} and
1450-1550 {\AA}, {\Lint}. This is because the non-ionizing UV photometry provides the
scaling for our SED model and {\Lint} subsequently defines the
intrinsic level of LyC flux. The difference between this intrinsic
flux and the observed $u$-band photometry will then define
{\fesc}. Given there is a high level of uncertainty regarding the specific star-formation
histories of high redshift galaxies, here we employ a non-parametric SED
fit using linear regression and BPASSv2.1 SED models \citep{eldridge17} rather
than constraining our models by assuming a fixed star-formation history.

For VUDS511227001 we first extract the 31 photometric observations for
from the COSMOS catalogs cover a rest wavelength range of $\sim$900 to $\sim$9600. We also
compile the filter transmission curves for all included filters. All photometric
fluxes and associated errors are then converted to $\mu$Jy and stored
in an input table along with the MOSFIRE spectroscopic redshift. In
our SED fitting, we exclude fluxes of any band with a rest wavelength shorter 
than 1216{\AA} as this probes the LyC/Ly$\alpha$ forest portion
of the specturm, thus the observed fluxes will also
depend on {\fesc} and/or $T_{\rm IGM}$. This leaves 20 photometric fluxes for our
SED fitting procedure. The SED fitting described here is primarily performed to
constrain the intrinsic LyC flux and E(B-V), which are then used to
independently estimate {\fesc} as described in Section
\ref{section:fescest}. 

Next, we prepare model photometric observations of BPASSv2.1 models,
which will be used to assess the goodness of fit of each
template. This is done by first converting the raw BPASS spectral
models from their provided units (ergs s$^{-1}$ {\AA}$^{-1}$ 10$^{-6}$
L$_{\odot}$) to $\mu$Jy $M_{\odot}^{-1}$ at a given redshift. We then
measure the weighted average flux per $M_{\odot}$ for each model in
each photometric band from the COSMOS field where the
weighting is given by the transmission curve of a given
filter. Ten sets of BPASSv2.1 spectra are provided at
metallicities between $z = 10^{-5}$ and $z = 0.014$ with each set containing 51 SSP models
in the age range log$_{10}$(age) = 6 - 11 years in bins of 0.1
dex. All metallicities are considered simultaneously while only the youngest
33 SSP ages are considered as older SEDs would be older than the age of
the universe at the redshift of VUDS511227001. Thus, at each metallicity we produce
a grid of 20$\times$33 photometric fluxes with each row giving the photometric fluxes at a
single age. Each individual grid is then stacked to produce a final grid
of 200$\times$33 photometric fluxes with each row representing an SED at a single
age and metallicity. In all cases we employ BPASSv2.1 models including binary
stellar evolution, a powerlaw initial mass function (IMF), and an
maximum stellar mass of 300 $M_{\odot}$ \citep[see][for more
details]{eldridge17}, similar to other works considering LyC emission
at high redshift \citep[e.g.][]{steidel18}. 

We include dust attenuation in our SED fitting employing a
\citet{reddy16} attenuation curve with
$R_{V} = 2.74$. Generally, known LyC emitting galaxies exhibit little
or no dust attenuation
\citep{vanzella10,shapley16,bian17,vanzella18,steidel18}, thus we test
models with E(B-V) values in the range 0.0 to 0.2 with $\Delta$
E(B-V) = 0.001. 

\begin{figure*}
  \includegraphics[width=\textwidth]{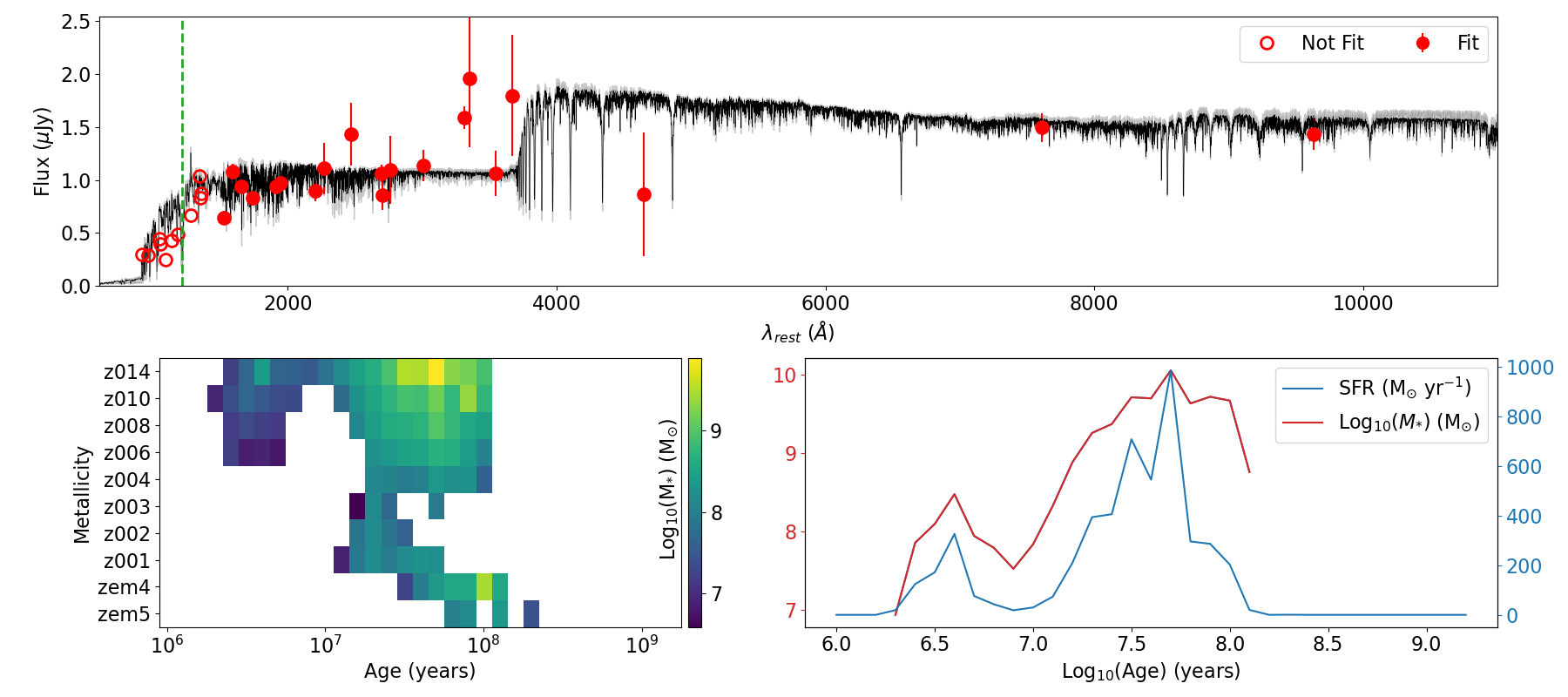}
    \caption{\textit{Top:} SED fitting results for VUDS511227001. The average
      SED across our 250 fits is shown in black with the 99$^{\rm th}$ percentile
      range shown as a shaded region. Solid red points show photometry included in
      the fitting procedure while open points show those that have been left out. These
      latter photometric observations are from filters that partially or entirely probe
      wavelength below Ly$\alpha$ (dashed vertical line) that are affected by IGM
      attenuation. \textit{Bottom left:} The average stellar mass of each of our 330
      single age, single metallicity BPASSv2.1 models across our 250 fits. The mass
      distribution represented here is typical of the majority of our fits. \textit{Bottom Right:}
      The average age vs stellar mass and star-formation history of our SED fits. Typical
      star-formation histories are dynamic and often characterised by a significant burst
      around 10$^{6.5}$ years ago peaking above 200 $M_{\odot}$ yr$^{-1}$. This recent
      burst is likely the primary contributor of LyC photons to the galaxy's SED.}
    \label{fig:sedresults}
  \end{figure*}  

With our photometric grid and dust curves in place, our SED fitting
procedure is performed. At each E(B-V) value we scale the
photometric fluxes of the dust free grid to match the expected fluxes
for our attenuation curve. We then use linear regression to
determine the best fitting linear combination of dust attenuated SSP
templates. Our linear regression is
achieved by minimising the cost function $J(\Theta,X,Y,\sigma_{Y})$
taken as the reduced $\chi^{2}$ value:
\begin{equation}\label{eq:cost}
  J(\Theta,X,Y,\sigma_{Y}) = \frac{1}{2m}\sum_{i=1}^{m}\frac{(h(\Theta,X)-Y)^{2}}{\sigma_{Y}^{2}}
\end{equation}
where $m$ is the number of photometric bands considered (20), $X$ is the
grid of model photometric fluxes, $Y$ are the observed fluxes for a
given object, and $\sigma_{Y}$ are the associated measurement errors
on the observed fluxes. Here the function $h(\Theta,X)$ describes a 1D
vector that represents the output SED where:
\begin{equation}\label{eq:h}
  h(\Theta,X)_{i} = \sum_{j=1}^{n}\Theta_{j} \times X_{j,i}
\end{equation}
where $n$=200 is the number of different age/metallicity BPASSv2.1 SSP models
considered. In this framework, and given our fluxes
are in $\mu$Jy per $M_{\odot}$, $\Theta$ can be seen as a vector
describing the amount of stellar mass attributed to each SSP model of a
given age with a length equal to the number of different aged models
being considered at a given redshift. As an example, assume $i$=0
corresponds to the $B$ band flux, thus $h(\Theta,X)_{0}$ represents
the linear combination of $B$ band fluxes across all SSP ages with
weighting given by the vector $\Theta$.

\begin{figure}
  \includegraphics[width=\columnwidth]{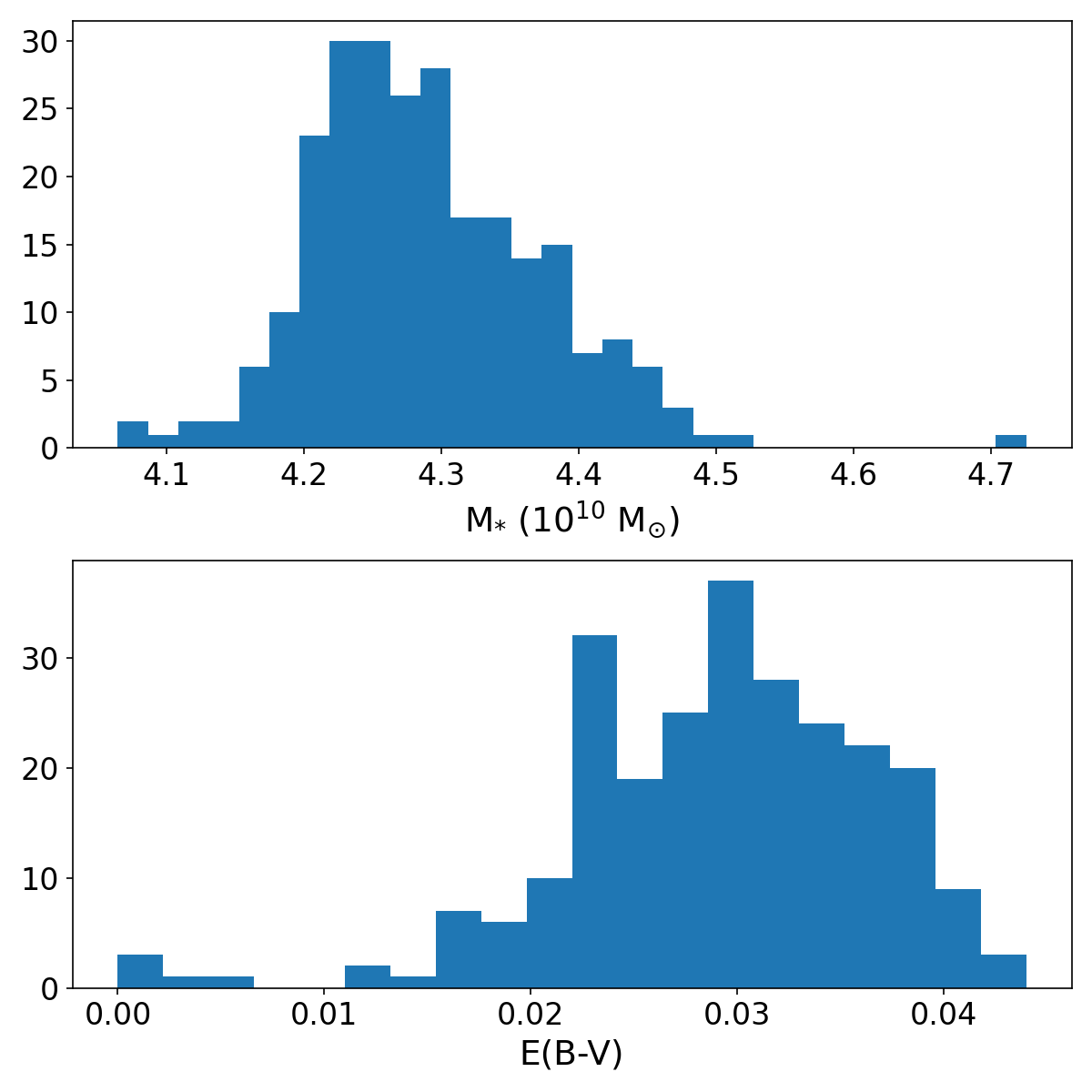}
    \caption{The distributions of stellar mass, top, and E(B-V),
    bottom, for the 250 SED models produced in our fitting procedure.
    Both stellar mass and E(B-V) are relatively consistent across the
    ensemble of models with the former found to be aroun 10$\times$10$^{4.3}$
    and the latter peaking around 0.03.}
    \label{fig:ebvmstar}
  \end{figure}

The resulting best fit SED for a single realisation of our fitting procedure
will also depend on the initial conditions due to the nature of the optimisation
algorithm. We thus perform the fitting 250 times using a randomly initialised
guess where the mass associated with each row is selected from a normal
distribution with both $\mu$ and $\sigma$ equal to 10$^{5}$ M$_{\odot}$. After
initialisation, any negative value is set to zero. Furthermore, the final
values are constrained between 0 and 10$^{10}$ M$_{\odot}$. We show the average
output SED, which includes dust attenuation, in the top panel of Figure \ref{fig:sedresults} with the shaded area
showing the 99$^{\rm th}$ percentile range among all 250 fits. The average mass
for each of the 330 single age stellar populations is shown in the bottom left
of Figure \ref{fig:sedresults}, which is representative of a typical fit result.
As shown in Figure \ref{fig:ebvmstar}, the resulting stellar mass falls in the
range 4$\times$10$^{10}$ to 4.8$\times$10$^{10}$ $M_{\odot}$. We show the average star-formation
history in the bottom right panel of Figure \ref{fig:sedresults}, which is found
to by quite dynamic. Typical SED fits exhibit a strong burst roughly 10$^{6.5}$
years ago peaking above 200 $M_{\odot}$ yr$^{-1}$, and it is likely that this
stellar population is the source of a majority of the LyC emission originating
in this galaxy. We also note that, although dust attenuation is typically relatively
low as can be seen in Figure \ref{fig:ebvmstar}, the intrinsic LyC emission from
SED fits with higher E(B-V) is necessarily larger, which will be reflected in the
estimated {\fesc} for such models.

It should be noted that the picture of star-formation histories for massive galaxies
at high redshift are highly uncertain. Although not presented here, we have also performed
SED fits using exponentially declining and constant star-formation histories, which result
in a similar spectral shape. Thus, although the physical parameters such as $M_{\odot}$
and metallicity may vary depending on the details of star-formation history, the intrinsic
LyC to UV flux, {\Lint}, is fairly consistent across all models. This value is one of the
key drivers of the resulting estimate of {\fesc}, meaning our results are not strongly
dependent on our choice of star-formation history.

\subsection{IGM Transmission}\label{section:IGMtrans}

In order to produce probability distribution functions (PDF) of
{\fesc} for VUDS 511227001 for each of the SED models described in
Section \ref{section:SEDmodels} we need to also sample the PDF of the IGM
transmission, {\TIGM}, for the $u$-band at that galaxy's
redshift. Here, we produce 10,000 IGM transmission curves using the
{\sc TAOIST-MC} code\footnote{available at
  https://github.com/robbassett/TAOIST\_MC} , based on the methods
described in \citet{inoue14} and \citet{steidel18}. 

A full description of the methodology for producing IGM transmission
curves can be found in \citet{bassett21}, however we briefly describe
the major details here. For each stochastically produced IGM
transmission curve we first generate a single realisation of possible
intervening HI absorption systems in redshift bins of $\Delta z = 5
\times 10^{-5}$ from $z=0$ to the redshift of VUDS 511227001,
$z=3.64$. Here, absorption systems are sampled from the ``IGM+CGM''
column density distribution function presented in Appendix B of
\citet{steidel18}. Next, for each absorption system, the transmission
to ionizing radiation as a function of wavelength is determined at the
particular redshift of that system. Finally, the cumulative IGM
transmission function is computed as the combined transmission of all
absorption systems from $z=0$ to 3.64. This process is repeated 10,000
times to produce our ensemble of IGM transmission curves.

\begin{figure}
\includegraphics[width=\columnwidth]{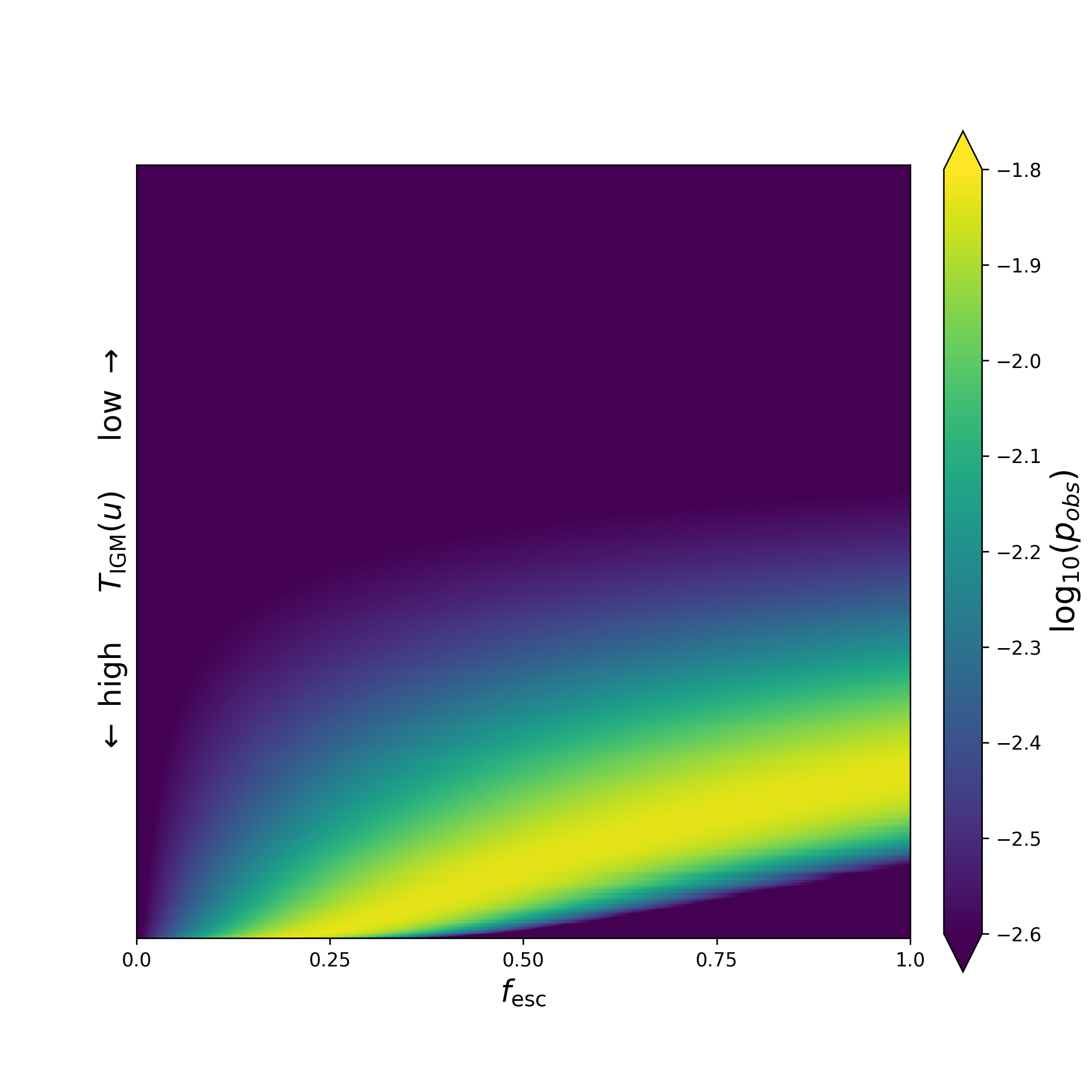}
  \caption{A 2D representation of the PDF of {\fesc} for 10,000 individual
    IGM sightlines for a single SED. Each row in this figure represents a single
    realisation of the IGM transmission curve where we have ordered
    the rows in decreasing $u$-band transmission, $T_{\rm IGM}(u)$, from bottom to
    top. As one might expect, the most probable value of {\fesc} for
    individual sightlines increases with decreasing IGM
    transmission. It can also be seen that for the majority of
    sightlines with relatively low transmission, the PDF is roughly
    flat with a value consistent with the probabilty for a $u$-band
    flux of 0. When computing the final {\fesc} PDF for each
    individual metallicity and age we determine the probability of a
    single {\fesc} value by taking the median of this figure along the
  $y$-axis.}
  \label{fig:pdf2d}
\end{figure}

\begin{figure}
\includegraphics[width=\columnwidth]{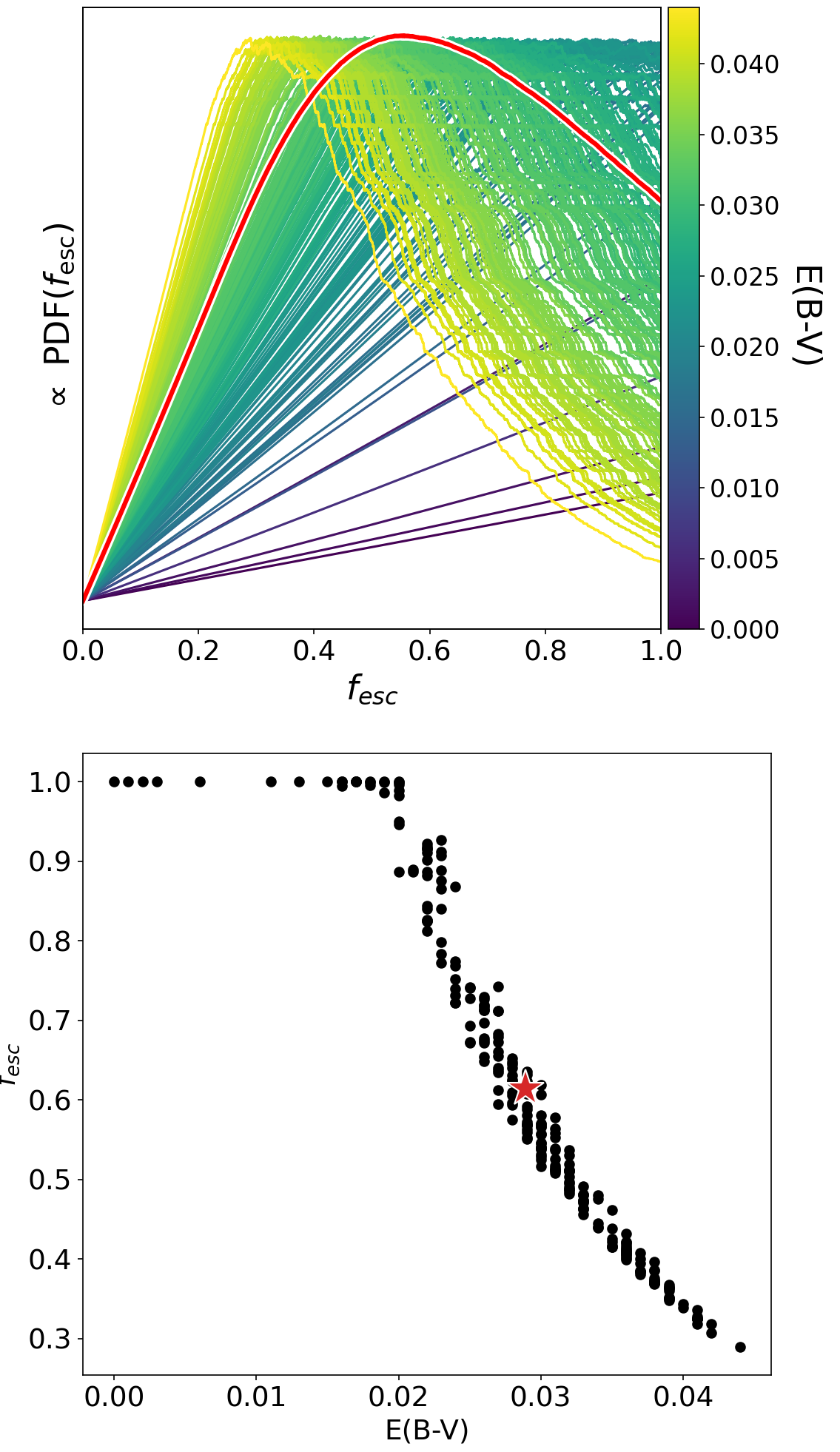}
  \caption{\textit{Top:} Individual {\fesc} PDFs produced as described in
  Section \ref{section:fescest} with colour indicating the E(B-V) of a
  particular model. The average PDF across all 250 models is shown in red,
  and it is from this average PDF that we compute the median value and
  uncertainty of {\fesc} for VUDS511227001. \textit{Bottom:} The relationship
  between the peak {\fesc} and E(B-V) for each of our models. As can be inferred
  from the top panel, we find a decrease in {\fesc} with increasing E(B-V). This
  is due to the fact that the intrinsic SED of a dust free model closely matches
  the observed photometry meaning a smaller difference between the intrinsic and
  observed LyC flux.}
  \label{fig:fescpdfs}
\end{figure}

\subsection{{\fesc} Estimates}\label{section:fescest}

\subsubsection{{\fesc} From Statistical Analysis}\label{section:fescpdf}

Having produced a series of SED models with similar goodness of fit and produced an ensemble of
IGM transmission curves for VUDS 511227001, we now describe our method
for determining the posterior PDF of {\fesc} for each model. The goal
here is to determine the probability that the observed $u$-band flux
and associated error is consistent with a given value of {\fesc}
considering the ensembles of input SED models and IGM transmission
curves described above. 

For each of our 250 SED fits we test all 10,000 IGM transmission curves,
as well as 10,000 {\fesc} values between 0 and 1. We
first multiply the intrinsic model spectrum (i.e. in the absence of
dust attenuation) with a given IGM transmission
curve, producing the SED+IGM model consistent with {\fesc} = 1.0. We
then apply {\fesc} values between 0 and 1 with $\Delta${\fesc} =
1$\times 10^{-3}$. Here, we simply assume a flat transmission value
for LyC photons such that all photons with $\lambda_{\rm rest} <
911.8$ {\AA} are attenuated by the same value, namely the current
{\fesc}. Next we measure the $u$-band flux of the SED+IGM+{\fesc}
model by taking its weighted average with weighting given by the
CLAUDS $u$-band transmission curve, which includes the combined
effects of optics, CCD quantum efficiency, and the average atmospheric
transmission at Mauna Kea. This value is taken as a single
mock observation. We then determine the probability, $P_{\rm obs}$, that a given mock
observation is consistent with the observation as the value of a
Gaussian function with $F(u _{0})$ and $\sigma(u)$ taken from the
observed flux and error, respectively, evaluated at the position of
the mock observation's $F(u)$.

This process is visualised for a single SED in Figure \ref{fig:pdf2d}
where each row represents a single IGM sightline and sightlines have
been ordered by decreasing $u$ band IGM transmission from bottom to top. For each
sightline the colorbar shows the correspondence between a given
{\fesc} value and $P_{\rm obs}$ where the highest value in a given row
gives the {\fesc} value most consistent with the observed
photometry. We note that for nearly half of our sightlines, the IGM
transmission is so low that the resulting $u$-band flux is $\sim$0 for
all {\fesc} meaning the value of $P_{\rm obs}$ is the same at all
{\fesc}. In contrast, for the highest transmission sightlines at high
{\fesc} the resulting $u$ flux is more than twice the observed value
meaning the combination of high transmission and high {\fesc} is less
consistent with the observations than a $u$ flux of 0 (see Section
\ref{section:results} for further discussion). 

After applying all 10$^{8}$
combinations of {\TIGM} and {\fesc}, the probability of any individual
value of {\fesc} is taken as the median across all 10,000
sightlines. This is equivalent to taking a one dimensional median of
Figure \ref{fig:pdf2d} along the y-axis.
In this way, we construct the PDF of {\fesc} for an
individual combination of SED age, E(B-V), and metallicity.

\begin{figure}
\includegraphics[width=\columnwidth]{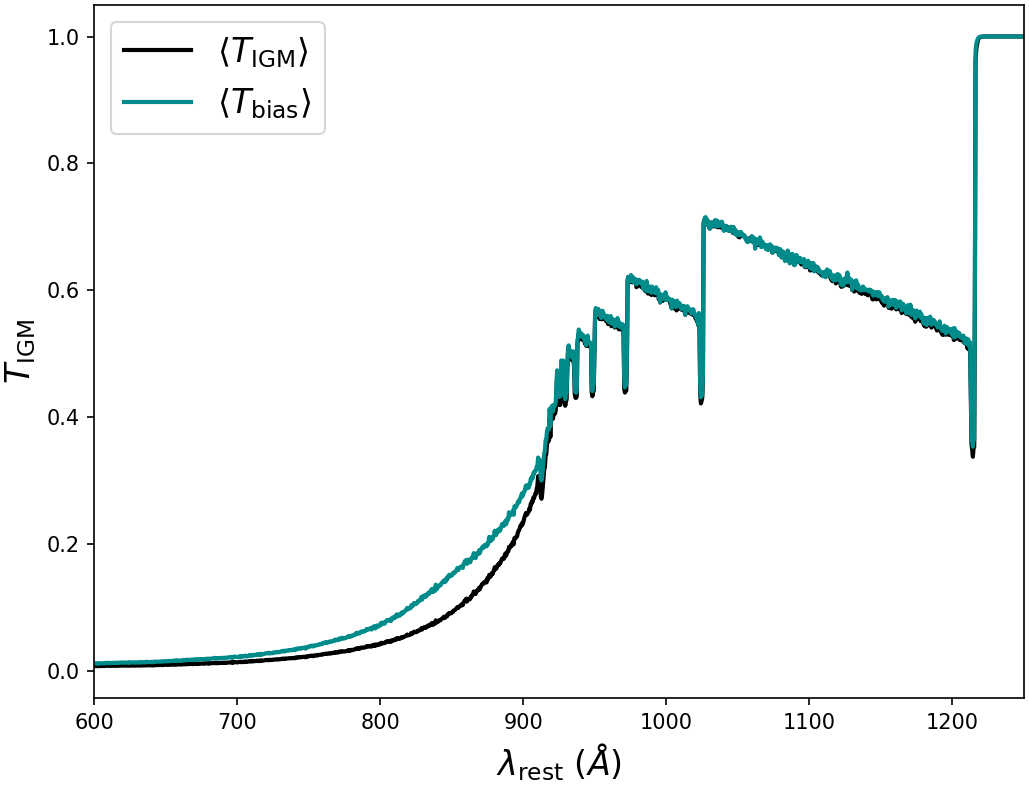}
  \caption{Shown in black is the average IGM transmission across
  all 10,000 simulated sightlines, while in cyan is shown the average
  biased transmission across all 250 SED models. For a single SED the
  biased IGM transmission is calculated as the weighted average of
  all 10,000 sightlines with weight given by the sum of the {\fesc} 
  PDF produced for each individual sightline (i.e. individual rows
  in Figure \ref{fig:pdf2d}). The average transmission in the $u$-band
  for the biased transmission is 0.03 higher than the unweighted average
  transmission.}
  \label{fig:Tbias}
\end{figure}

\begin{figure}
  \includegraphics[width=\columnwidth]{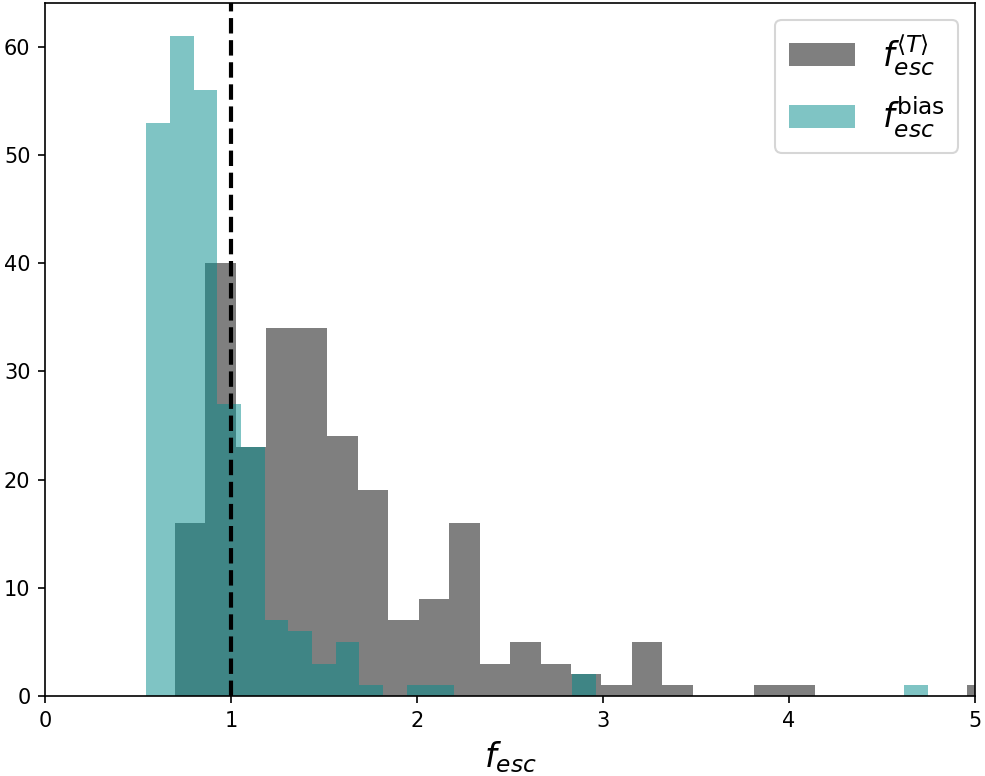}
    \caption{Histograms of {\fTm} and {\fbias}, the {\fesc} values
    estimated by assuming the average and the biased IGM transmission
    at the redshift of VUDS511227001. In the case of {\fTm}, the vast
    majority of {\fesc} values have unphysical values larger than 1.
    For {\fbias}, the majority of values, as well as the average value,
    fall just below {\fesc} = 1. Regardless, {\fbias} is significantly
    larger than {\fpdf} considering the modest average $T_{\rm bias}$ of 0.03.}
    \label{fig:fescnormal}
  \end{figure}


This process is repeated across the entire ensemble of
SED models produced in Section \ref{section:SEDmodels}, producing a broad ensemble of
{\fesc} PDFs. We show all 250 {\fesc} PDFs in the top panel of Figure \ref{fig:fescpdfs} with
the colour indicating the E(B-V) value for a given model. We see the peak
location of the PDF increases with decreasing E(B-V). This relationship is
shown in the bottom panel of Figure \ref{fig:fescpdfs} which shows a strong
decrease from {\fesc} = 1.0 for models with E(B-V) $<$ 0.02 down to {\fesc}
$\simeq$ 0.3 for our most attenuated model. This reflects the fact that
the attenuated model is fit to the observed photometry while the intrinsic
model has a higher flux with the difference increasing with E(B-V).

We also produce a single {\fesc} PDF to consider
the \textit{entire} ensemble of models. This is achieved by
taking the weighted average of all 250 PDFs with the weight given 
by the peak sum of a given PDF. This weighting scheme gives a lower
weight to SEDs with low E(B-V) as the peak, found at {\fesc} = 1.0,
is significantly below the peak value for models with maximums below
{\fesc} = 1.0. We show the resulting
average PDF in red in the top panel of Figure \ref{fig:fescpdfs}.

It is from the weighted average {\fesc} PDF that we estimate the most
probably value of {\fesc} for VUDS511227001. We also provide assymetric
errors with the error on each side given by the difference between the
most probable {\fesc} and the 15.9$^{\rm th}$ and 84.1$^{\rm th}$
percentiles, roughly equivalent to the 1$\sigma$ spread in the case of
a Gaussian distribution. We also
compare the value of {\fesc} computed using the method described here
with {\fesc} computed for the same target using alternative methods
(see below),
thus for the remainder of this work this value of {\fesc} will be
presented as {\fpdf}. The resulting value of {\fpdf} from this process
for VUDS 511227001 is 0.51$^{+0.33}_{-0.34}$. 

\subsubsection{{\fesc} From Average IGM Transmission}\label{section:fescavg}

A common method of estimating {\fesc} for LyC detected galaxies is to
assume $T_{\rm IGM}$ equal to the average value at a given
redshift \citep{bian17,fletcher18,naidu18}. This method is more appropriately applied to large samples
of LyC detections as in this case differences in IGM sightline from
galaxy to galaxy average out and an average value of {\fesc} can be
estimated. To use the average IGM transmission in this way, however,
requires large samples of LyC detections at roughly fixed redshift. As
such samples are not existent currently, the average transmission
method has been applied to single galaxies in some previous
works. While we do not advocate such an application, we 
calculate {\fesc} in this way for VUDS 511227001 for illustrative
purposes. We note that, because we assume a CGM contribution to the 
attenuation of LyC photons \citep[following][]{steidel18} the {\fesc}
values estimated in this manner will be higher than if we had omitted
this contribution \citep[e.g.][]{inoue08,inoue14}.

Following a number of previous works
\citep[e.g.][]{steidel01,bassett19,mestric20} we calculate {\fesc}
from the average IGM transmission in the following manner. We first
calculate the so-called relative {\fesc}, $f_{\rm esc}^{\rm rel}$ as:
\begin{equation}\label{eq:ftm}
  f_{\rm esc}^{\rm rel} = 
  \frac{(F_{LyC}/F_{1500})_{\rm obs}}{(L_{LyC}/L_{1500})_{\rm int}}
  \times \frac{1}{\langle T_{\rm IGM} \rangle}
\end{equation}
where $(F_{LyC}/F_{1500})_{\rm obs}$ is the observed LyC to 1500
{\AA} flux ratio, $(L_{LyC}/L_{1500})_{\rm int}$ is the SED dependent,
  intrinsic LyC to 1500 {\AA} luminosity ratio, and {\Tm}
  is the transmission of the IGM to LyC radiation measured here for
  the CLAUDS $u$ band at the redshift of VUDS 511227001 of
  $z=3.64$. We then convert $f_{\rm esc}^{\rm rel}$ to the absolute
  {\fesc}, $f_{\rm esc}^{\rm abs}$ as:
\begin{equation}
  f_{\rm esc}^{\rm abs} = f_{\rm esc}^{\rm rel} \times 10^{-0.4(k_{1500}E(B-V))}
\end{equation}
where $k_{1500}$ is the reddening at 1500 {\AA} for our chosen
\citet{reddy16} attenuation curve and E(B-V) is the best fitting value
as calculated for each of our SED models in Section
\ref{section:SEDmodels}. For IGM transmission curves produced with TAOIST-IGM
at $z=3.641$, we find a value of {\Tm} for the $u$-band of 0.04$\pm$0.01.

Previous work has shown that for LyC surveys, there is
an expected bias such that the IGM transmission along sightlines
towards LyC detected galaxies is higher than the mean value
\citep{riverathorsen19,bassett21}. This bias, which we call $T_{\rm bias}$, is
expected to vary depending on a variety of factors including
observational depth, brightness of detected sources, and the intrinsic
SED shape of observed sources (in particular the LyC to 1500 {\AA}
flux ratio). While \citet{bassett21} primarily focused on quantifying
$T_{\rm bias}$ for samples of LyC detections, one can also compute the
expected $T_{\rm bias}$ for individual detections. We note that,
although IGM transmission is not an additive quantity, we prefer to
define it as such for convenience. Thus, $T_{\rm bias}$ is defined as
the difference between the average IGM transmission in sightlines in
which a given galaxy is more likely to be observed and the average IGM
transmission of all sightlines at a given redshift \citep[see][for
more details]{bassett21}. 

\begin{figure*}
\includegraphics[width=\textwidth]{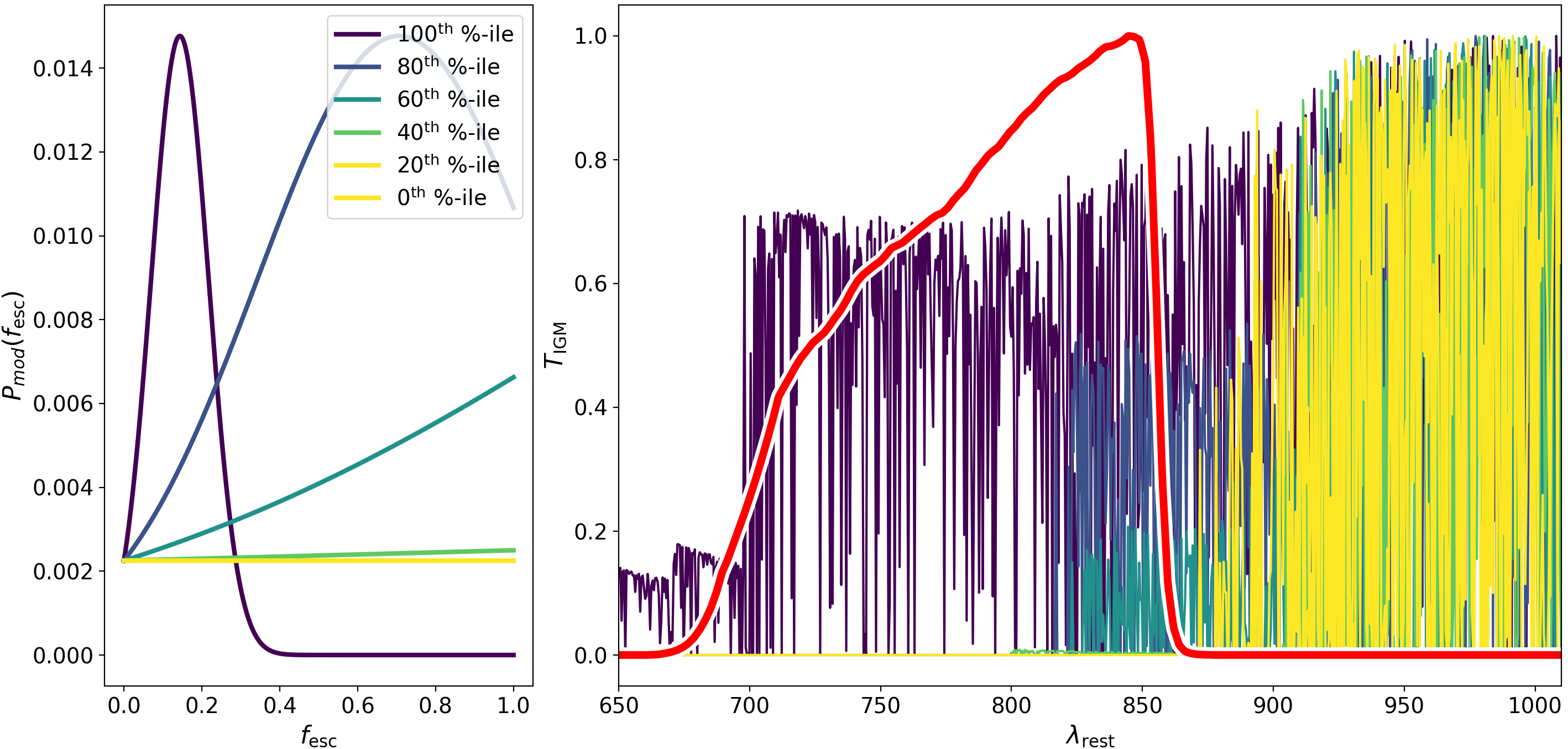}
  \caption{\textit{Left:} The output probability distributions of
    {\fesc} for individual IGM sightlines covering a range in
    {\Tm}($u$). Here we have ordered all 10,000 sightlines by
    {\Tm}($u$) and display the output PDF for percentiles between 0
    and 100. Here, the 100$^{\rm th}$ percentile represents the most
    transparent sightline in our analysis. \textit{Right:} For
    reference the IGM transmission functions for each of the
    PDFs shown in the left panel are displayed. We also show the
    trasmission of the $u$ band, scaled to have a peak value of 1, in red.}
  \label{fig:pdfindiv}
\end{figure*}

We demonstrate the bias in IGM transmission for VUDS 511227001 in
Figure \ref{fig:Tbias}. The black curve in each shows the
unweighted mean transmission curve at $z=3.64$ while coloured curves
show the weighted average considering the observed photometry. Here
the weighting for each combination of SED and IGM transmission is
determined by summing the resulting {\fesc} PDF for that combination 
(i.e. summing a single row in Figure \ref{fig:pdf2d}).
The weighted average IGM transmission using this weigthing scheme is 
shown in Figure \ref{fig:Tbias} in cyan. To measure {\Tbias} we take the
weighted average of both curves in Figure \ref{fig:Tbias} with weights
given by the CLAUDS $u$-band transmission curve and subtract the resulting
value for the unweighted mean transmission from the biased mean transmission.
$T_{\rm bias}$ for the CLAUDS $u$-band for our 250 SED models
is found to be 0.03$\pm$0.01 giving {\Tm}$+T_{\rm bias}$ 0.07$\pm$0.02. 

Having computed $T_{\rm bias}$ for each SED model, we can then
computed a bias corrected value of {\fesc} as:
\begin{equation}\label{eq:fbias}
  f_{\rm esc}^{\rm bias} = 
\frac{(F_{LyC}/F_{1500})_{\rm obs}}{(L_{LyC}/L_{1500})_{\rm int}}
  \times \frac{1}{\langle T_{\rm IGM} \rangle + T_{\rm bias}}
\end{equation}
Although the level of $T_{\rm bias}$ expected for VUDS 511227001
appears negligible at $\sim$0.03, estimating {\fesc} using Equation
\ref{eq:fbias} results in an appreciable difference when compared to
Equation \ref{eq:ftm}. 

For both {\fTm} and {\fbias} the report values in Section \ref{section:results}
are taken as the mean across our 250 SED models. We also provide an 
assymetric error as the 15.9$^{\rm th}$ and 84.1$^{\rm th}$ percentiles
of the same 250 values.
Given we are providing three different methods of calculating {\fesc}
we distinguish these as {\fpdf} for the probability based method from
Section \ref{section:fescpdf}, {\fTm} for values calculated with Equation
\ref{eq:ftm}, and {\fbias} for values calculated with Equation
\ref{eq:fbias}. We reiterate that we consider {\fpdf} to be the most
rigorous as it considers the sightline-to-sightline variation in
$T_{\rm IGM}$, which typically has a more complex wavelength
dependence than smooth curves produced by averaging over ensembles of
sightlines. Our results suggest that these complex variations between
individual sightlines can result in significant differences in calculated
{\fesc} values and should be taken into account when interpreting
detections of LyC radiation from individual galaxies.

\section{Results}\label{section:results}

In this Section we have a closer look at the PDFs of {\fesc} produced
in Section \ref{section:fescest} for VUDS 511227001. We begin by 
comparing the composite {\fesc} PDFs shown in
Figure \ref{fig:fescpdfs}, which take into consideration all 10,000
IGM sightlines, with {\fesc} PDFs produced for individual sightlines
at a range of $T_{\rm IGM}$. Next, we will compare the final {\fpdf}
values with the two more traditional {\fesc} estimates, {\fTm} and
{\fbias}, produced from averaged IGM transmission curves. For clarity
we state simply here that the resulting {\fesc} values of our three
methods calculated for VUDS 511227001 are {\fpdf} =
0.51$^{+0.33}_{-0.34}$, {\fTm} = 1.40$^{+0.80}_{-0.42}$, 
and {\fbias} = 0.82$^{+0.33}_{-0.16}$. These
values consider all 250 SED models produced in our analysis in a
probabilistic manner (as described above).

In Figure \ref{fig:pdfindiv} we show {\fesc} PDFs of individual IGM
for a single SED model. Shown are the 0$^{\rm th}$, 20$^{\rm th}$, 40$^{\rm th}$,
60$^{\rm th}$, 80$^{\rm th}$, and 100$^{\rm th}$ percentiles in IGM
transmission for the CLAUDS $u$-band. We can
see for sightlines with high IGM transmission the PDF is roughly
Gaussian with the peak indicating the particular {\fesc} value exactly
matching the observed $u$-band flux for a given
sightline. Additionally, we see that the probability that 0 flux is
consistent with the observed flux is $\sim$0.002, thus all sightlines
overlap here for {\fpdf} = 0. We also note that, as can be seen in
Figure \ref{fig:pdf2d}, the majority of sightlines have a roughly flat
PDF as the LyC flux is $\sim$0, thus having a probability of
$\sim$0.002, regardless of {\fpdf}. In contrast, for sightlines with
high IGM transmission, when {\fpdf} is large the resulting $u$ band
flux is more than twice the observed value. This results in a
$P_{\rm obs}$ value for such sightlines significantly lower than 0.002
(the value for $F(u) = 0$) meaning that the combination of high IGM
transmission and high {\fesc} is less consistent with the observed
value than a $u$ flux of 0. This is precisely the reason that the
median $P_{\rm obs}$ at high {\fesc} is lower than for intermediate
values, resulting in a PDF that peaks below 1. 

We also wish to point out that the final PDFs for {\fpdf} can vary
significantly from the PDFs for any individual sightlines. The most
transparent sightline, for instance, is roughly Gaussian and narrowly
peaked with the most probable {\fesc} value at $\sim$0.19,
significantly lower than the most probable value considering the full
ensemble of IGM sightlines. Given any single galaxy exists along a
single IGM sightline, were it possible to know the true transmission
toward which VUDS 511227001 is observed the PDF would likely collapse
to a more well defined value. Further, if this sightline were of
particularly high transmission, the most probable value would exist at
a relatively low probability considering the ensemble PDFs shown in
Figure \ref{fig:fescpdfs}. This point must be kept in mind when
consdering {\fesc} values for individual galaxies as it is likely that
larger samples of LyC detected galaxies are required to fully
understand the role of star-forming galaxies in driving reionization.

\begin{table}
  \vspace{2mm}
  \begin{center}
  \begin{tabular}{ c c c c }
    \hline\hline
    {\fpdf} & {\fTm} & {\Tbias} & {\fbias} \\
    \hline
    0.51$^{+0.33}_{-0.34}$ & 1.40$^{+0.80}_{-0.42}$ & 0.032$^{+0.009}_{-0.010}$ & 0.82$^{+0.33}_{-0.16}$ \\
    \hline
  \end{tabular}
  \caption{Median {\fesc} values for each of our different estimators. Assymetric represent the 15.9$^{\rm th}$
    and 84.1$^{\rm th}$ percentile range, equivalent to 1$\sigma$ for a Gaussian distribution.}
  \label{table:fescpdf}
  \end{center}
\end{table}

We next compare {\fpdf} with the more traditional value of {\fTm} and
the value {\fbias}, which takes into account the bias towards
detecting galaxies from sightlines with nonzero IGM transmission. We
can see in both cases that these two values result in significantly
higher {\fesc} estimates when compared to {\fpdf}, although {\fbias}
is within the upper bounds of the 1$\sigma$ spread in
{\fpdf}. Considering the final value that takes into account all
metallicities and ages, bottom row of Table \ref{table:fescpdf}
(noting again this is most closely associated with our $Z_{*}$ = 0.14
model), we find {\fpdf} = 0.51$^{+0.34}_{-0.33}$, 
{\fTm} = 1.40$^{+0.80}_{-0.42}$, and {\fbias} = 0.82$^{+0.33}_{-0.16}$. We note that
estimates of {\fesc} for VUDS 511227001 have been performed previously
by \citet[][their galaxy ID 368]{mestric20} who provide a range of values for different
{\Lint}. Our value of 1.40$^{+0.80}_{-0.42}$ is consistent with the values presented in
\citet{mestric20} of {\fesc} $\gtrsim 0.3-0.93$. 

Considering the three calculations of {\fesc} presented here we find
that the values of {\fTm} and {\fbias} here are larger than {\fpdf} by 
0.88 and 0.30, respectively. It is also worth noting that {\fbias} is
lower than {\fTm} by 0.58, a significant decrease even though the
average level of $T_{\rm bias}$ we calculate is small at
$\sim$0.03. This further highlights the importance of carefully
considering the fact that LyC detections, by definition, are
incompatible with IGM sightlines with very low or 0 transmission,
which are often included in the calculation of {\Tm}. 

What then is the driver of the significant difference between {\fpdf}
and both {\fTm} and {\fbias}? Ultimately, the determination of the
probability of a given value of {\fesc} for an individual combination
of IGM sightline and intrinsic SED shape is defined by the resulting
$u$-band flux. Given that we use the same ensembles 
SED models for all {\fesc} determinations, we are left with the
differing treatment of the IGM transmission. Clear differences can be seen by
comparing average IGM transmission curves in Figure \ref{fig:Tbias}
with individual curves from Figure \ref{fig:pdfindiv} with the former
characterised by a smooth transition from high to low $T_{\rm IGM}$
with decreasing wavelength and the latter by sharp, stochastic drops
in $T_{\rm IGM}$. Put another way, the smooth, average transmission
curves commonly seen in the literature
\citep[e.g.][]{inoue14,steidel18,bassett21} are not representative of
individual sightlines and the stochastic nature of such individual
transmission curves results in a posterior distribution of $u$-band
fluxes that is not well captured assuming {\Tm} even when correcting
for {\Tbias}. Some previous works \citep[e.g.][]{shapley16,vanzella16}
have used ensembles of IGM transmission curves in their estimates of
{\fesc} resulting in similar distributions of {\Tm} as seen here. Thus,
the small novelty introduced in this work is the consideration of the
observed flux and uncertainty to apply observational probabilities to
individual sightlines resulting in a posterior {\fesc} distribution
peaking below {\fesc} = 1.0.

We end this Section by reiterating that we consider this rigorous
method of determining {\fpdf} to be most appropriate for determining
{\fesc} for individual galaxies. This method takes into account the
current best understanding of the probability of LyC photons (at
wavelengths where the $u$-band is sensitive in the observed frame)
encountering high column density neutral hydrogen along any individual
line of sight. We repeat that the ensemble PDF produced this way may
not be representative of the true PDF given any single galaxy has only
one IGM transmission curve. What we have shown, however, is that
methods considering only the mean IGM transmission, or even the biased
average transmission \citep{bassett21}, are likely not providing
reasonable estimates of individual {\fesc} values. Given the large
uncertainties for individual estimates, however, it is most likely
that statistically significant samples of LyC detections will be
required to truly understand the role of star-forming galaxies in
reioinzation. Furthermore, it is also possible that the smooth, average
transmission curves used to calculate {\fTm} and {\fbias} are indeed
appropriate when applied to larger samples of galaxies at roughly
fixed redshift.

\section{Summary \& Conclusions}\label{section:conclusions}
 
In this paper we begin with a 
sample of eight galaxies selected as LyC emitting candidates based on CLAUDS
$u$-band detection. All of our targets were selected with
prior redshift estimates at $z > 3.4$ such that the detected $u$-band
flux can be attributed only to LyC photons with $\lambda_{\rm rest} <
911.8$ {\AA}. Previous redshift estimates for 7/8 of our sample come
from ZFOURGE photometric analysis with the remaining galaxy selected
from VUDS, which provides a secure spectroscopic redshift. Thus, we begin our analysis by
measuring the spectroscopic redshifts of our ZFOURGE selected targets
from [OIII] $\lambda$5007 {\AA} 
(along with [OIII] $\lambda$4959 {\AA} and H$\beta$, where detected)
emission lines in Keck MOSFIRE spectroscopy. 

Our first result is that all seven galaxies selected from ZFOURGE
found to have overestimated photometric redshifts. In all cases, MOSFIRE
spectroscopic redshifts are found at $z < 3.4$. This means that the $u$-band 
detections are contaminated by Lyman $\alpha$ forest photons, 
preventing us from providing meaningful constraints on the LyC escape
fraction {\fesc}. Our systematic selection of galaxies with overestimated
photometric redshifts
may suggest that our requirement of a $u$-band detection has resulted in a
bias such that we preferentially select galaxies with overestimated
photometric redshifts rather than clean detections of LyC emission (see Appendix \ref{section:zoverest}).
This result provides a useful warning to other projects searching for LyC
detections based only on photometric estimates of galaxy
redshifts. Whether or not this is a general problem or related to the
particular photometric bands and methods of the ZFOURGE survey or to
any associated selection biases, however, is yet to be seen.

The final galaxy in our sample, VUDS 511227001 had a spectroscopic
redshift estimate prior to our MOSFIRE observations. Our detections of
[OIII] $\lambda$5007 {\AA} confirm the redshift of this target to be
$z=3.64$, thus its $u$-band detection results purely from LyC
photons. We reiterate that this galaxy has been analysed in the
context of LyC escape previously by \citet[][VIMOS
spectroscopy]{marchi17} and \citet[][also CLAUDS $u$ band]{mestric20}.
For this target we have performed a rigorous statistical
analysis to determine the PDF of {\fesc}. In our analysis we have
produced 10,000 IGM transmission functions to $z=3.64$ and employed an
ensemble of 250 SED models constructed from BPASSv2.1 templates (see \ref{section:SEDmodels}).
Our SED models are non-parametric, but are have a characteristic mass of 
4.25$\times$10$^{10}$ $M_{\odot}$ and exhibit bursty star-formation histories.
Critically, for determination of LyC escape, these
models cover a range of {\Lint} between 0.02 and 0.15. 
We then produce PDFs of {\fesc} for each SED in a probablistic manner
by applying all 10,000 IGM transmission functions across values of
{\fesc} between 0 and 1 then taking into consideration the observed
$u$-band flux and error of VUDS 5112270001 (for more details see
Section \ref{section:fescavg}). Finally, we determine a final PDF of
{\fesc} across all 250 models as the weighted average of each
individual PDF with weights given by the probability that a given SED
matches the observed ZFOURGE photometry.

The final PDF of {\fesc} for VUDS 511227001 is shown in Figure
\ref{fig:fescpdfs} in red, and we refer to the most probable value as
{\fpdf}. The resulting value for this target is 0.51$_{-0.34}^{+0.33}$
where the asymmetric errors represent the $1\sigma$ range of our
PDF. We also calculate two alternative values of {\fesc} using the
mean IGM transmission, {\Tm}, and mean transmission included the
expected observational bias {\Tm}+{\Tbias}
\citep[see][]{bassett21}. These values, which we refer to as {\fTm}
and {\fbias} are found to be 1.40$^{+0.80}_{-0.42}$ and 0.82$^{+0.33}_{-0.16}$, respectively, noting that
the expected level of bias in {\Tm} for this target is $\sim$0.03. The
value calculated for {\fTm} is consistent within errors with the previous estimate
of \citep{mestric20} who use a similar method. We
postulate that the large differences between these estimates of
{\fesc} result from the fact that individual IGM transmission
functions are characterised by sharp drops not well represented by the
smoothly declining functions seen in averaged IGM transmission
curves. The complex shape of individual sightlines convolved with the
SED model shape and the transmission of the $u$-band result in a
distribution of model $u$-band fluxes not well represented by the
simplified calculations of {\fTm} and {\fbias}. 

Ultimately, we also question the value of individual measurements of
{\fesc}, even using a rigorous method such as the one presented
here. Any observation of a galaxy represents a single IGM
transmission, and we have shown that knowing the exact form of this
transmission can result in a PDF not well represented by the PDF we
produce by marginalising over 10,000 such sightlines. Thus, it is
likely that a real understanding of the role of star-forming galaxies
in cosmic reionization will require much larger samples of LyC
detected galaxies, and in such cases average transmission curves
become more appropriate.

\section*{Data Availability}

IGM transmission functions used in this work is produced primarily using publicly
available codes found at https://github.com/robbassett as well as
publicly available galaxy SED models from the BPASS collaboration
\citep{eldridge17}. Observational data from Keck is available from the
Keck science archive and CLAUDS data will be released publicly in the
near future (likely late 2021).

\section*{Acknowledgements}

This research was conducted by the
Australian Research Council Centre of Excellence for All Sky
Astrophysics in 3 Dimensions (ASTRO 3D), through project
number CE170100013. Some of these data were obtained and processed as part of the CFHT Large Area
U-band Deep Survey (CLAUDS), which is a collaboration between
astronomers from Canada, France, and China described in Sawicki et
al. (2019, [MNRAS 489, 5202]).  CLAUDS is based on observations
obtained with MegaPrime/ MegaCam, a joint project of CFHT and
CEA/DAPNIA, at the CFHT which is operated by the National Research
Council (NRC) of Canada, the Institut National des Science de
l’Univers of the Centre National de la Recherche Scientifique (CNRS)
of France, and the University of Hawaii. CLAUDS uses data obtained in
part through the Telescope Access Program (TAP), which has been funded
by the National Astronomical Observatories, Chinese Academy of
Sciences, and the Special Fund for Astronomy from the Ministry of
Finance of China. CLAUDS uses data products from TERAPIX and the
Canadian Astronomy Data Centre (CADC) and was carried out using
resources from Compute Canada and Canadian Advanced Network For
Astrophysical Research (CANFAR). (Some of) The data presented herein
were obtained at the W. M. Keck Observatory, which is operated as a
scientific partnership among the California Institute of Technology,
the University of California and the National Aeronautics and Space
Administration. The Observatory was made possible by the generous
financial support of the W. M. Keck Foundation. 
The authors wish to recognize and acknowledge the very significant
cultural role and reverence that the summit of Maunakea has always had
within the indigenous Hawaiian community.  We are most fortunate to
have the opportunity to conduct observations from this mountain. Results
presented in this work have made extensive use of the python3
programming language \citep{python3} and, in particular, the authors wish to
acknowledge the the numpy \citep{numpy}, matplotlib \citep{matplotlib},
and scipy \citep{scipy} packages. MR
and LP acknowledge support from HST programs 15100 and 15647. Support
for Program numbers 15100 and 15647 were provided by NASA through a
grant from the Space Telescope Science Institute, which is operated by
the Association of Universities for Research in Astronomy,
Incorporated, under NASA contract NAS5-26555. MS acknowledges support
from the Natural Sciences and Engineering Research Council (NSERC) of Canada.




\bibliographystyle{mnras}
\bibliography{refs} 



\appendix

\section{Systematically Overestimated Photometric Redshift for $u$-band
  Selected Galaxies}\label{section:zoverest}

Here we take a closer look at the systematic
overestimate of ZFOURGE photometric redshifts for all seven galaxies
selected based on those redshifts. In all cases, this overestimate
was large enough that the true redshift, based on [OIII] 5007 {\AA}
detections, was found to be below $z \lesssim 3.4$, the redshift limit for
CLAUDS-$u$ to cleanly sample the LyC emission. As we have stated, this finding ultimately
means that estimates of {\fesc} from the $u$-band detections for these
galaxies are almost entirely unconstrained.

These targets provide
a useful cautionary result regarding LyC candidate selection based on
the combination of photometric redshift and expected LyC detection. The question we
pose here is: is it possible that our sample selection methodology has
induced a bias such that we are more likely to select galaxies with
overestimated photometric redshifts from the ZFOURGE survey? Indeed, the
photometric redshift accuracy quoted by the ZFOURGE team based on
spectroscopic follow up is $\lesssim$2\% with roughly equal mix of
under and overestimated photometric redshift \citep[e.g.][]{straatman16}. We reiterate that this
statement is based primarily on lower redshift galaxies where the ZFOURGE
medium band filters directly probe the Balmer break, which is not the
case at the redshift of our sample. This would suggest that our selection of 7/7
galaxies with overestimated photometric redshifts is statistically
unlikely from random selection alone.

\begin{figure}
\includegraphics[width=\columnwidth]{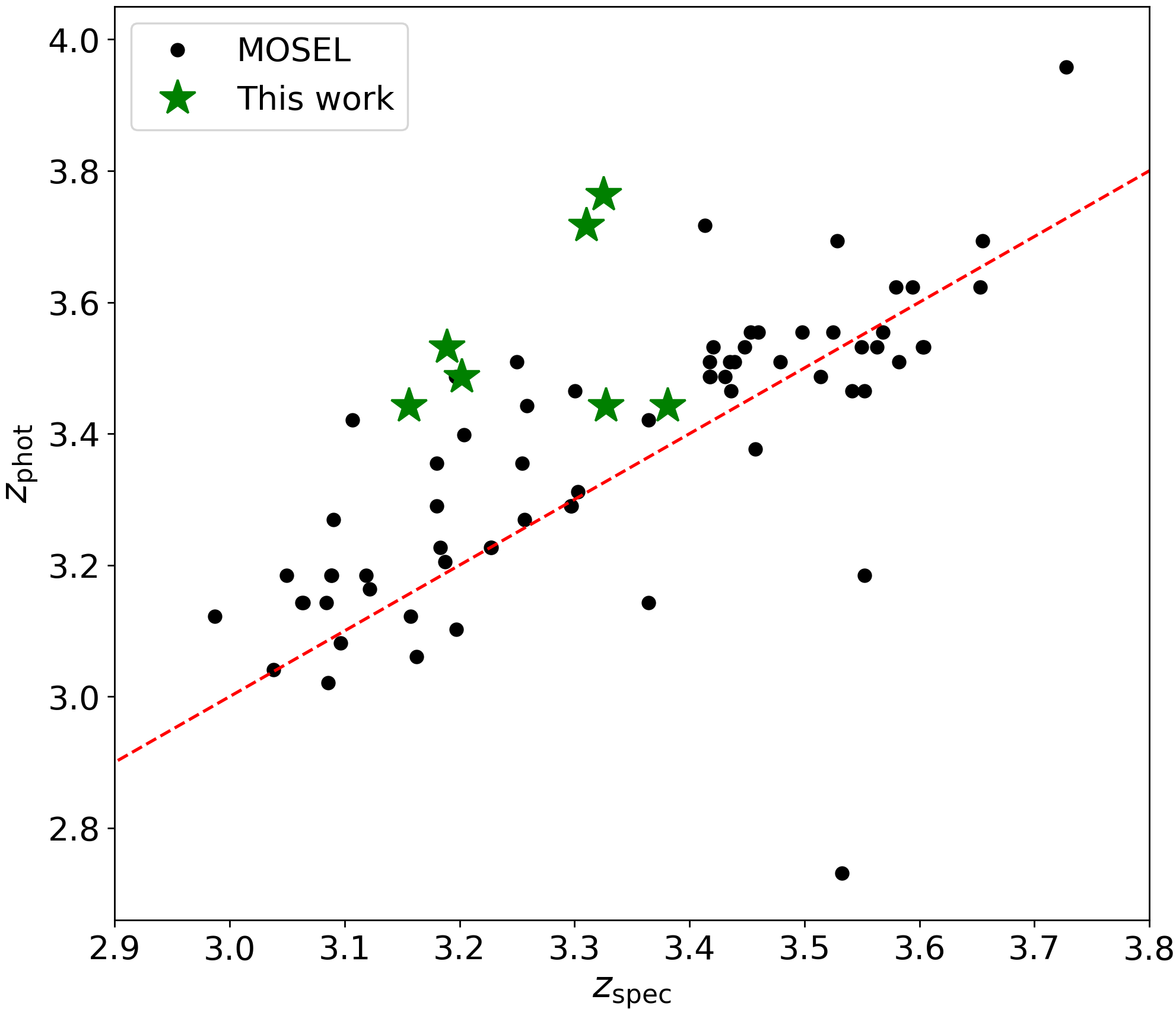}
  \caption{Photometric versus spectroscopic redshift measurements for
    our sample and a sample of MOSEL \citep{tran20} galaxies at a similar
    redshift. Here we show that 7/7 of our ZFOURGE targets are found
    to have overestimated photometric redshift estimates from the
    ZFOURGE catalog. In comparison, 47/70 of the MOSEL galaxies have
    similarly overestimated photometric redshifts. If we take MOSEL as
  a parent sample, we find a probability of 0.062 that we have
  selected 7 galaxies with overestimated photometric redshifts by
  chance. We postulate instead that our selection criteria,
  in particular the requiredment of a detection in the CLAUDS $u$-band, have
  resulted in a bias towards selecting galaxies with overestimated
  photometric redshifts.}
  \label{fig:zcomp}
\end{figure}

To test for a statistical effect, we have obtained a sample of 70
ZFOURGE galaxies at $z > 3$ 
having spectroscopic follow up observations taken from the
Multi-Object Spectroscopic Emission Line
survey \citep[MOSEL][]{tran20}. We show in Figure \ref{fig:zcomp} a
comparison between photometric and spectroscopic redshifts for both MOSEL
galaxies (black circles) and our sample (green stars). The sample of
MOSEL galaxies shown here represents the 
largest sample of ZFOURGE galaxies at $z > 3$ with measured $z_{\rm
  spec}$ currently known, thus we will use this sample to determine
the likelihood that spectroscopic redshift vs photometric redshift 
for our sample is consistent with a random selection from ZFOURGE. 

From Figure \ref{fig:zcomp} it can be seen that, similar to our
sample, there is a tendency for ZFOURGE photometric redshift to be slightly
overestimated at $z > 3$. Indeed, 47/70 galaxies in the MOSEL sample
have overestimated photometric redshift. To determine if our selection of
7/7 overestimated photometric redshift is consistent with a random sampling
of MOSEL galaxies we employ binomial statistics: either galaxies have
underestimated photometric redshift or they don't. From this simple test, we
can calculate the probability of selecting 7 out of 7
galaxies with overestimated photometric redshift among $z > 3$ galaxies in
ZFOURGE as $P(k;n,p) = (47/70)^{7}$, giving 0.062. Furthermore, we
note that the 5/7 of our ZFOURGE targets are found near the upper
limit of $z_{\rm phot}$ - $z_{\rm spec}$ of the MOSEL sample meaning
the true probability of selecting galaxies with such a large
photometric redshift overestimate is even less likely than the simple
binary statistics estimate presented here. Thus, we consider
it improbable that
we have selected 7 galaxies with overestimated photometric redshift by
chance. 

An alternative explanation is a bias
induced by our selection requirements: $z_{\rm phot} > 3.4$, lack of
close companions in space-based imaging, and a clear detection in the
CLAUDS $u$-band. For a given galaxy
to satisfy the final requirement of $u$-band detection, at least one of three
things must be true. Either the galaxy truly is a high redshift LyC
emitter, the detection is contaminated by a low redshift interloper
unresolved even in HST imaging, or the photometric redshift is overestimated and
the $u$-band detection is contaminated by brighter Ly$\alpha$ forest
emission. Given the high spatial resolution of HST imaging, the second
possibility is unlikely. The liklihood of finding true LyC emitters is
still somewhat uncertain with detections rates in recent surveys in
the 5-20\% rate \citep{fletcher18,steidel18,mestric20}, with possible
dependence on selection type \citep[i.e. LBG vs LAE,
e.g.][]{bassett21}.  Based on the fact that $>$2/3 of
MOSEL galaxies have overestimated photometric redshifts, the liklihood
of a biased selection of galaxies with overestimated photometric
redshifts is more likely than selection of true LyC emitters
with our ZFOURGE+CLAUDS selection. Thus, we conclude that
our selection aimed at identifying LyC 
emitters at $z > 3.4$ based solely on photometric redshifts has
induced a selection bias that undermines our efforts at measuring
{\fesc} at high $z$. 


\bsp	
\label{lastpage}
\end{document}